\documentclass[prd,aps]{revtex4}
\usepackage{graphicx}

\begin{document}
\input psfig
\pssilent

\title{Canonical quantization of general relativity: the last
18 years in a nutshell}

\author{Jorge Pullin}

\affiliation{
Department of Physics and Astronomy, Louisiana State
University, 202 Nicholson Hall, Baton Rouge, LA 70803}

\date{September 3rd 2002}

\begin{abstract}
This is a summary of the lectures presented at the Xth Brazilian
school on cosmology and gravitation. The style of the text is that of
a lightly written descriptive summary of ideas with almost no
formulas, with pointers to the literature. We hope this style can
encourage new people to take a look into these results. We discuss the
variables that Ashtekar introduced 18 years ago that gave rise to new
momentum in this field, the loop representation, spin networks,
measures in the space of connections modulo gauge transformations, the
Hamiltonian constraint, application to cosmology and the connection
with potentially observable effects in gamma-ray bursts and conclude
with a discussion of consistent discretizations of general relativity
on the lattice.
\end{abstract}
\maketitle

\section{Introduction: where is string theory in these lectures?}

The words ``quantum gravity'' have become associated by physicists
over the years with ``difficult problem''. Attempts to quantize
general relativity started almost immediately after quantum mechanics
was establishes (see the article by Rovelli \cite{rohistory} for a
concise history of the field). Among the people involved are the most
stellar names in physics. Indeed, one should expect problems when
attempting to apply the rules of quantum mechanics to general
relativity. Although general relativity was developed before quantum
mechanics, the latter was introduced in the context of Newtonian
physics. Already the incorporation of special relativity required some
effort and in fact, the introduction of quantum field theory, which in
many ways is an extension of quantum mechanics. General relativity
however, is a much more radical revision of physics than special
relativity. It is a theory of space-time itself as opposed to a theory
of entities living in a spacetime. Quantum mechanics was firmly based
on the latter viewpoint. This key element separates general relativity
from almost all other physics theories. The invariance of the theory
under coordinate redefinitions, which is more clearly viewed as
invariance of the variables under diffeomorphisms, is not present in
any other significant physical theory. In fact, we have only learnt
how to apply the rules of quantum mechanics to theories invariant
under diffeomorphisms relatively recently \cite{Birmingham:1991ty}. 
And the theories in question,
like BF theory or Chern--Simons theory, are remarkably simpler than
general relativity. These theories are only superficially field
theories, in that they are described in terms of fields, but the true
degrees of freedom of the theories are finite in number. They are, in
fact, mechanical systems instead of field theories.  Solutions of the
equations of motion are given by fields that are ``trivial'', the only
non-trivialities coming from possible topological features of the
manifolds the theories live on. Once one realizes this fact, it should
not be surprising that their quantization becomes 
relatively straightforward.

There is a strong sociological element involved in quantum gravity
as well. After the many successes of quantum field theory following
World War II, it could only be expected that the application of the
same powerful techniques to general relativity should finally conquer
the problem. But this was not so. The application of perturbation theory
to general relativity taught many interesting lessons, consolidating
the ideas of gauge and ghosts. But it ultimately appeared to fail.
General relativity appears to be perturbatively non-renormalizable.
The practitioners of quantum field theory became so discouraged by
this fact, that they adopted the point of view that general relativity
should be abandoned as a physical theory. It is not that the successes
of the theory explaining the classical world are in question. It is
the fundamental nature of the theory. In this point of view general
relativity would play the role of an effective theory. The Lagrangian
and field equations of general relativity should be viewed  as, 
for instance, those of the Navier--Stokes theory of fluids. 
A highly successful and useful theory, but not one that anyone would
care to quantize, for instance, to describe a quantum fluid. Just 
like in the case of quantum fluids one quantizes a theory underlying
the Navier--Stokes one, one would quantize a theory underlying 
general relativity. A theory that reproduces general relativity 
only in certain regimes, but is richer in other regimes. Another
analogy that comes to mind is the Fermi theory of weak interactions,
non-renormalizable and just a low energy manifestation of a richer
theory, the theory of weak interactions. This point of view
led to the development of supersymmetric theories, supergravity,
Kaluza-Klein theories, superstrings and M-theory.

Is this point of view the only one? Strictly speaking, the answer is 
yes. General relativity only describes gravity, and therefore a 
richer theory should come into play in order to have a unified 
picture of all interactions, so general relativity indeed should be
the limit of a larger theory. But even if one ignores all other
interactions, are we completely sure that general relativity
cannot be quantized? This appears as an academic question. After all,
if we know we need a larger theory, why bother with determining
if general relativity can be quantized? The reason this question
is, in the view of some people, not of purely academic interest,
is that general relativity has features that we would all desire
in a unified description of nature. Most notably, the fact that
space-time is dynamical and invariant under diffeomorphisms.
In a sense, general relativity is perhaps the simplest theory
incorporating these features with non-trivial content. Lessons 
learnt from attempting to quantize it should therefore be very
valuable at the time of quantizing a richer theory. These are the
reasons propelling a small but non-trivial (see \cite{Rovelli:1997qj}
for some statistics) minority of physicists
to study the quantization of general relativity.

But isn't the fact that the theory is non-renormalizable an
indictment of this program? How could one quantize such a theory?
To understand this we need to separate an intrinsic question
(is the theory quantizable or not) from a procedural question
(can we quantize it using perturbation theory). These two
questions are different. In fact, we know of examples of 
theories that admit a quantum description and that we do not
know how to treat perturbatively. De Witt's group 
\cite{deLyra:1992up} studied sigma models
that have this feature. But more striking is the example of
general relativity in $2+1$ dimensions. In dimensions lower than
four, the Einstein equations just state that the metric is flat.
General relativity in such a situation is only an apparent field
theory, since the only solution to the equations is ``constant''.
One can have degrees of freedom if the topology of the manifold
is non-trivial. Yet, when perturbative quantization of general
relativity in $2+1$ dimensions was attempted, the theory appeared
to be non-renormalizable more or less in the same way the four
dimensional version was. It was only when Witten 
\cite{Witten:1988hc}
noticed that 
one should be able to perform a non-perturbative quantization,
and carried it out, that people realized one could find ways
to treat the theory perturbatively \cite{Deser:xu}.

The general relativity in $2+1$ dimensions example exhibits in 
a dramatic way the pitfalls expecting anyone attempting to
quantize these kinds of theories. The lesson is that the
symmetries of these theories are far more elaborate than
usual. In the case of $2+1$ gravity, the symmetry group
is so large that the theory is rendered trivial by it. 
Quantization schemes that do not take this into account,
fail. In $2+1$ dimensions, unravelling the symmetry was
easy because one has full control of the theory. The general
exact solution of the equations of motion is not only known
in closed form, but a good intuitive handle on its meaning
is available. Nothing like this occurs in $3+1$ dimensions.
Learning how to gain a comparable handle is the task at
hand. It is obviously a difficult task. We will never have
the general solution of the Einstein equations in four
dimensions in closed form. That may prevent us from ever
getting the kind of intuitive handle on the theory that
is needed in order to quantize it.

The non-perturbative quantization of gravity was pioneered
by DeWitt in the 60's, following the early efforts of Dirac
and Bergmann. An immediate problem that was encountered is
that the kind of variables in terms of which gravity is usually
described, is very different from the ones used in the 
successful quantum field theories of particle physics. In 
the latter, the fundamental variable is a connection.
This made many of the techniques that had been developed for
handling particle physics theory not applicable to general
relativity. A change in this situation occurred when 
Ashtekar introduced a formulation of general relativity
in terms of a connection that had a very elegant and simple
canonical structure. In fact, the theory resembled a Yang--Mills
theory and opened the possibility of introducing the techniques
so successfully used in that context to general relativity.
These lectures will give glimpses onto some of the results
that have arisen ever since.

\section{Organization and coverage}

These notes are based on lectures. Due to the finite lecture time
(further compressed by a two day plane delay!) and lack of expertise
of the author in some areas it was not possible to cover many
topics. A big, broad topic I missed is spin foam approaches to the
path integral.  This will be covered soon by a forthcoming review
paper by Perez \cite{Perez}. I will not discuss the beautiful results
on black hole entropy. These are of  great importance, since they are
precise calculations that do not shy away from taking into account the
full dynamics of the theory. The paper by Ashtekar et al.
\cite{Ashtekar:2000eq} has references to all the early
literature. Very recent work by Varadarajan \cite{Varadarajan:2002ht}
and others \cite{Ashtekar:2001xp}, showing a connection between the
loop representation and more traditional Fock pictures and the work of
Thiemann's group \cite{Thiemann:2002vj} on semi-classical states could
not be covered.  The beauty of these results requires a level of
detail that was not possible in the format of the lectures. Finally,
although the notes attempt to guide a newcomer to the literature, they
have not been prepared carefully enough as to attempt to be a
comprehensive review. Loll
\cite{Loll:1998aj}, Rovelli
\cite{Rovelli:1997yv} and Thiemann
\cite{Thiemann:2001yy}Living Reviews
articles covering lattice approaches, loop quantization and 
canonical quantum gravity respectively. Carlip
\cite{Carlip:2001wq} 
has a superb review on quantum gravity in general, giving the
essentials of all approaches.  For a lighter reading, Smolin's \cite{3road}
recent book covers several aspects of quantum gravity. Detailed
discussion of the early results are found in Ashtekar's books \cite{Asbook}, other
topics can also be seen in the book we wrote with Gambini \cite{Gapubook}.
Baez and Muniain have an introductory book to knot theory with applications
to gravity \cite{Baez:sj}.

\section{Canonical quantization}

Canonical quantization is the oldest and most conservative
approach to quantization. It demands one to control the theory
well, not allowing to bypass several detailed questions, namely
what is the space of states of the theory, what is the inner
product, what is observable. Every physicist has performed a
few canonical quantizations in courses on quantum mechanics.
To canonically quantize one roughly
follows the following steps: a) one picks a Poisson algebra of
classical quantities that is large enough to span the physics
of interest (in ordinary mechanics $q$ and $p$, for instance);
b) one represents these quantities as operators acting on a 
space of wavefunctions and the Poisson algebra as an algebra
of commutators; c) if the theory has constraints, that is,
quantities that vanish identically classically, one has to 
impose that they vanish quantum mechanically as operators;
d) an inner product has to be introduced on the space of
wavefunctions that are annihilated by the constraints; 
e) Predictions for the expectation values of observables 
(quantities that have vanishing Poisson brackets with the
constraints) can be worked out.
Notice that several of these steps involve choices. For 
instance, there is no unique way to choose an inner product,
or to choose a certain set of classical quantities to be
promoted to operators.

For general relativity one has to start by casting the theory in a
canonical form. This was done by Dirac and Bergmann in the 50's and
60's (for a more modern discussion see \cite{Asbook}) . The
fundamental canonical variable is the metric of a spatial surface
$q_{ab}$ (people normally use $q$ instead of $g$ to avoid confusion
with the space-time metric). Its canonically conjugate momenta is
usually denoted $\tilde{\pi}^{ab}$ and the tilde denotes that it is a
density, as momenta usually are. The momenta are closely related to
the extrinsic curvature, which is also closely related to the time
derivative of the three metric. The theory has four constraints.
These are relationships among the variables at a given instant
of time.
Three of them form a vector and are called the ``vector'' or
diffeomorphism constraint. When one has constraints in canonical
theories, it is due to the presence of symmetries. The diffeomorphism
constraint can be shown to be associated with the invariance of
general relativity under spatial diffeomorphisms.  The remaining
constraint is associated with the invariance of general relativity
under diffeomorphisms off the spatial surface.  It is usually called
the scalar or Hamiltonian constraint.  Unfortunately, the canonical
treatment breaks the symmetry between space and time in general
relativity and the resulting algebra of constraints is not the algebra
of four diffeomorphisms, the Hamiltonian constraint is singled out and
behaves differently. The algebra of constraints is closed in the sense
that Poisson brackets of constraints are proportional to constraints,
but the Poisson bracket of two Hamiltonian constraints is proportional
to a diffeomorphism constraint through a function of the canonical
variables. This we will see, will cause difficulties at the time of
quantization.

If one performs a Legendre transform one finds that the Hamiltonian
of general relativity is just a combination of the above mentioned
constraints. That is, the Hamiltonian of the theory vanishes. This
is due to the fact that the notion of time introduced in order to
set up the canonical theory is a fiducial, arbitrarily introduced
time. The canonical formalism ``knows'' that relativity does not
really single out a preferred time and responds back by saying that
the Hamiltonian associated with any artificial time vanishes.
Physical time can only be retrieved in the canonical theory through
an elaborate process with many difficulties (Kucha\v{r}'s \cite{ku} article
contains a detailed discussion of the ``problem of time'')

One can attempt a canonical quantization by considering wavefunctions
of the metric $\psi(q)$. One can represent the metric as a multiplicative
operator and its canonically conjugate momentum as a functional 
derivative. One can then attempt to promote the constraints to 
operatorial equations. The diffeomorphism constraint, which is linear
in the momenta, is relatively simple to implement. It implies that
the wavefunctions are really only functions of the diffeomorphism 
invariant content of $q$ and not of $q$ itself. This is natural and
elegant, but is also problematic: we do not know how to code in a
simple way the diffeomorphism invariant information of $q$. Therefore
the solution to constraint presented is natural but also quite formal;
we cannot write it or handle it in an explicit way. A worse situation
arises when one considers the Hamiltonian constraint. The latter is a
non-polynomial function of the canonical variables that requires 
regularization. Most regularizations used in particle physics depend
on the presence of a background (c-number) metric, which we do not
have available in quantum gravity. No satisfactory treatment of 
the Hamiltonian constraint in this context has ever been found.

Worse, the lack of control on the space of functions considered
also implies that we do not know any kind of useful inner product
to be introduced. 

Finally, in the last step we were supposed to compute expectation
values of the observables (see \cite{Rovelli:2001my} for references
on the observables problem) of the theory. Observables have to be
quantities that have vanishing Poisson brackets with the constraints.
This implies they are invariants under the symmetries of the theory
and that as quantum expressions they will act upon the space of
physical states (solutions to the constraints) in such a way as
to keep us within that space. Unfortunately no such quantities are
known for general relativity in a generic situation. If one 
reduces the theory by introducing additional symmetries, 
sometimes observables can be found. For 
instance in cosmological models or if the space-times considered
are asymptotically flat. What is happening here is that since the
Hamiltonian of the theory is a combination of constraints, finding
observables is tantamount to finding the constants of motion of
the theory. But the constants of motion can be seen as 
re-expressing the initial conditions for a given solution of the
equations of motion as functions of phase space. This, of course,
requires solving the equations of motion, something we cannot
do for general relativity in closed form, unless we have 
symmetries present. This has suggested the
possibility that the problem could perhaps be tackled in an
approximate form. Progress has been recently made on this issue
in the new variable context as well \cite{GaPulambda}.

\section{The new variables}

The introduction of Ashtekar's new variables generated the new
momentum that has invigorated the field in the last 18 years.
For pedagogical reasons, the new variables are best introduced in a two
stage process. The first stage is to use, instead of the metric of
space as a fundamental variable, a set of triads $\tilde{E}^a_i$ (the
tilde denotes a density weight, introduced for convenience, $a$ is a
spatial index and $i$ labels the three frame fields). People had
considered using the triad as a canonical variable. The description
closest to the notation used in these days is given by Barbero
\cite{Barbero:1994ap}. Extra
constraints arise since the theory is now invariant under frame field
transformations (rotations) as well. The Hamiltonian is still a
complicated non-polynomial function of the canonical variables. So the
introduction of triads per se is not too helpful. The real
breakthrough was the realization by Ashtekar that the Sen 
\cite{Sen:qb} connection
could be used as a canonically conjugate momentum to the
triad. Usually the canonically conjugate momentum considered for the
triad is proportional to the extrinsic curvature. The Sen connection
adds a piece given by the spin connection of the triad. Actually the
sum of these two terms can be done while multiplying the extrinsic
curvature times a constant (this constant is called the Immirzi parameter), 
yielding a one-parameter family of
possible canonically conjugate variables (all members of the family
are related by a canonical transformation).  We call this the
generalized Sen connection.  If one rewrites the constraints in terms
of the triads and the generalized Sen connection several things
happen.  The set of constraints introduced due to the symmetry of the
theory under triad rotations now takes the form of ``divergence of the
triad equal zero''. The divergence is taken with respect to the
generalized Sen connection.  If one writes the connection as $A$ and
the triads as $E$ the resulting equation is exactly a Gauss law $D_a
\tilde{E}^a_i=0$, like the one present in $SO(3)$ Yang--Mills theory
(the $SO(3)$ arises due to the symmetry under rotations of the triads)
.  The diffeomorphism constraint takes a form that resembles a
Poynting vector $\tilde{E}^a_i F_{ab}^i=0$. This is nice, since the
latter is clearly associated with the momentum of the fields and fields
without net linear momentum are the only ones invariant under diffeomorphisms.
Finally the Hamiltonian constraint still is a complicated
non-Polynomial function of the variables. However, if one chooses the
Immirzi parameter equal to the imaginary unit (this is if one
considers a Lorentzian signature space-time, for an Euclidean
signature the Immirzi parameter should be chosen equal to one), the
non-polynomialities cancel out. One is left with a Hamiltonian
constraint that takes a simple, polynomial (in fact at most quadratic)
form in terms of the canonical variables 
$\tilde{E}^a_i \tilde{E}^b_j F_{ab}^i=0$.

Another appealing aspect is that written in terms of these new variables,
general relativity appears as a Yang--Mills theory with a set of 
extra constraints (and with a different Hamiltonian). This opened the
possibility of introducing in general relativity tools that were used
in Yang--Mills theory for its quantization. One of these tools is the
use of loops.

\section{Loops and the loop representation}

We could now attempt a canonical quantization of the theory we just
discussed. One could pick wavefunctions that are functionals of the
connection $\psi[A]$ just like in Yang--Mills theory and promote the
connection and the triad to canonically conjugate operators. Notice
that this is already quite a departure from the traditional
quantization where one took functionals of the metric. One now has to
impose the constraints as operator equations. The Gauss law as an
operator just demands that the wavefunctions be $SO(3)$ invariant
(gauge invariant in the Yang--Mills language). 

An interesting set of gauge invariant functionals of the connection
is given by considering the trace of the holonomy of the connection
along a loop. 
\begin{equation}
W_\gamma(A)={\rm Tr}\left(P \exp\oint dy^a A_a\right).
\end{equation}
These quantities are called ``Wilson loops''. A very
attractive feature is that these quantities constitute a basis
for all gauge invariant functions \cite{Gi}. That is, given any gauge invariant
function of a connection it can in principle be expressed as a linear
combination of Wilson loops based on different loops. The coefficients
of this expansion therefore contain all the gauge invariant information
of the wavefunction. Therefore whenever we think of a gauge invariant
functional of a connection $\psi[A]$ one can alternatively think of 
the coefficients $\psi(\gamma)$ of its expansion in the Wilson loop
basis ($\gamma$ is a loop). Representing the functions in this way
is what is known as the ``loop representation''. In this representation
wavefunctions are functions of loops and operators are geometric 
operators that act on loops. Such representation was first proposed
for Yang--Mills theory by Gambini and Trias \cite{GaTr86} and for general relativity
in terms of the new variables by Rovelli and Smolin \cite{RoSm88}.

A caveat is that the basis provided by Wilson loops is really an
overcomplete basis. The coefficients in the expansion $\psi(\gamma)$
are therefore constrained by certain relations, known as the Mandelstam
identities \cite{Ma79}. For a function of a loop to be admissible as a wavefunction
in the loop representation, it should satisfy these identities. 
This can be challenging to achieve. As we will see a solution to
this problem was eventually found in terms of spin networks.

Since one is automatically considering gauge invariant functions only
when one works in the loop representation, Gauss law identically 
vanishes. We are therefore left with the diffeomorphism and Hamiltonian
constraints only. The beauty of the loop representation lies in the
natural action the diffeomorphism constraint acquires in this 
representation. The diffeomorphism constraint acts on wavefunctions
by shifting infinitesimally the loop. Therefore it is immediate to
solve the diffeomorphism constraint. One simply has to consider
wavefunctions of loops such that they are invariant under deformations
of the loops. Such functions are studied by the branch of mathematics
called knot theory and are known as knot invariants. We therefore
see that we can solve the diffeomorphism constraint in terms of a set
of functions on which there is a lot of mathematical knowledge.

A further surprise that the loop representation yielded was that it
appeared to also help in solving the Hamiltonian constraint
\cite{JaSm}. In retrospect, this result appears as of quite limited
importance, but it provided a quite significant boost of interest in 
the subject at the time it
was found, so we will review it here. Let us go back briefly to the
connection representation.  Suppose we want to promote the Hamiltonian
constraint to a quantum operator. We choose a factor ordering such
that the triads are to the right. One needs to regularize the
operator. Let us choose a simple minded point splitting, putting the
two triads and the curvature at slightly different points. The triads
operating on the Wilson loop produce a result that is proportional to
the tangent to the loop at the point where they act (one can see this
simply by noting that it is the only vector present at that point).
That means that the two functional derivatives, viewed as a tensor,
produce as a result a symmetric tensor, since the result is
proportional to a vector times itself. This tensor is contracted with
$F_{ab}$, which is antisymmetric. Therefore the result
vanishes. Notice that this result does not depend on the details of
the regularization.  This result was first noticed by Jacobson and
Smolin \cite{JaSm}. One caveat is that for it to be true, the loops in question
have to be smooth. If a loop is not smooth, it contains points where
there is more than one tangent and the constraint is not automatically
zero, since the two functional derivatives could be proportional to
different vectors and do not yield a symmetric tensor anymore.

If we now go back to the loop representation, the previous result
suggests that if we consider knot invariants $\psi(\gamma)$ that have
support on smooth loops only, one would appear to solve all the
constraints of quantum gravity. This result by Rovelli and Smolin \cite{RoSm88}
generated a lot of excitement. But there are problems in attempting to
solve the constraints in such a way. First of all, generic knot
invariants with support on smooth loops only fail to satisfy the
Mandelstam identities. This can be fixed by using spin networks as we
will see later. But more importantly, smooth loops appear to be too
simple to carry interesting physics.  As we will see, operators like
the volume of space vanish identically unless one has
intersections. Therefore this space of solutions of the constraints is
very likely a ``degenerate'' subspace that is not large enough to do
meaningful physics. It was however, of great historical importance.

Another early result of interest in the loop representation was based
on an observation due to Kodama \cite{Ko}. The observation is that if one
considers the exponential of the integral over space of the
Chern--Simons form built from the Sen connection $\exp\left(k S_{CS}\right)
=\exp\left(k \left[{\rm
Tr}\int A\wedge \partial A+ 2/3 A\wedge A\wedge A\right]\right)$, it
automatically solves the Hamiltonian constraint in the connection
representation if one introduces a cosmological constant. The
cosmological constant produces an extra terms in the Hamiltonian of
the form $\Lambda/6 \tilde{E}^a_i\tilde{E}^b_j\tilde{E}^c_k
\epsilon^{ijk} \epsilon_{abc}$. To see that the Kodama state
solves the constraint we only need to know that when one acts 
with $\delta/\delta A_a^i$ on it, one gets something proportional
to $\epsilon^{abc} F_{bc}^i$; this one can see through a calculation.
To put it in simpler terms, for this state ``$\tilde{E} \sim \tilde{B}$,''
that is the triad is proportional to the magnetic field built from the
Sen connection. Therefore the two terms in the Hamiltonian constraint,
which can be schematically written as ``$\hat{E}\hat{E}\hat{B}
+\Lambda/6 \hat{E}\hat{E}\hat{E}$'' can be made to cancel each other
by a choice of the constant $k$ in the Kodama state.

The Kodama state has been found to have connection, in the
cosmological context, with the Hartle--Hawking and Vilenkin vacua
\cite{Ko}. Of interest as well is its expression in the loop
representation. To transform this state to the loop representation we
wish to find its expansion in the basis of Wilson loops, that is, the
coefficients,
\begin{equation}
\psi(\gamma)=\int DA \exp\left(k S_{CS}\right) W_\gamma(A) 
\end{equation}
This integral has been extensively studied in a different context,
that of Chern--Simons theory \cite{Wi89}. That is, consider a theory whose 
action is $S_{CS}$. Then the integral can be viewed as the 
computation of the expectation value of the Wilson loop in a 
Chern--Simons theory. The result is a function of a loop that
is diffeomorphism invariant, that is, it is a knot invariant.
This invariant is the Kauffman bracket, which is closely related
to the Jones polynomial, a knot invariant of great interest in
the mathematical literature. Since this invariant is the transform
of a state that is annihilated by the Hamiltonian constraint in the
connection representation, it should be annihilated by the Hamiltonian
constraint in the loop representation. This has been checked 
explicitly \cite{BrGaPuprl}. It is remarkable that this invariant from the mathematical
literature, which was developed in completely independent fashion from
any idea related to the gravitational field, manages to solve the 
quantum Einstein equations.

\section{Formal developments}

As we discussed in the previous section, the early years after the
introduction of the new variables and the loop representation were
ripe with intriguing and promising results, 
that appeared to be a 
significant step forward in the construction of a canonical theory
of quantum gravity. However, many of the results were of formal
nature only. There was not a good control on the space of states
that one was operating on. Here we will mention some formal
developments that took place in the mid 90's that helped gain better
control on the calculations being performed.

\subsection{Spin networks}

As we discussed before, Wilson loops are an overcomplete basis for
gauge invariant functions. They are constrained by a set of nonlinear
identities called Mandelstam identities. The simplest such identity
comes from the following identity of $SU(2)$ matrices in the 
fundamental representation $({\rm Tr}A)({\rm Tr}B)={\rm Tr}AB+{\rm Tr}AB^{-1}$.
In terms of Wilson loops this would read $W_\gamma W_\eta= W_{\gamma\circ\eta}
+W_{\gamma\circ\eta^{-1}}$ with $\circ$ denoting loop composition. 
Now, it is clear that these identities stem from the fact that we 
are working in the fundamental representation of $SU(2)$ to
construct the Wilson loops. One does not need to do so. Consider
a diagram like the one in the figure. 
\begin{figure}[h]
\includegraphics*[height=5cm]{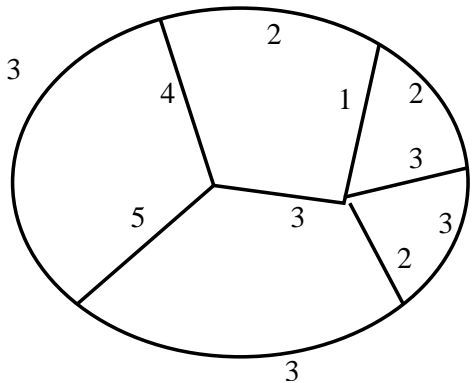}
\caption{A spin network.}
\end{figure}

One could construct holonomies along all the links in the diagram, in
principle, with different representations of $SU(2)$ on each link (the
representations are labeled by an integer). At each of the vertices
one would use invariant tensors in the group to ``tie up'' the
holonomies in such a way as to have at the end of the process a gauge
invariant quantity. Such a quantity is a natural generalization of the
Wilson loop. The diagrams like the one in the figure are called spin
networks. Since the invariants so constructed do not depend on any
particular representation, there are no relations between them as when
we were working in the fundamental representation only.  Therefore the
Mandelstam identities are automatically solved.  This was first
realized by Rovelli and Smolin \cite{Rovelli:1995ac}. Spin networks also
allow to do calculations in a natural and simple way, as we will see
in the following sections.

\subsection{Measures of integration}

Calculations like the ones needed to transform states to the loop
representation require a measure of integration in the space of
connections modulo gauge transformations. Such integrations are
also of interest to construct inner products. These are really
functional integrals in spaces of infinite dimensionality. 
There is little experience on how to construct such integrals.
Ashtekar and Lewandowski \cite{AsLe} among others have pioneered the 
construction of measures of integration in such spaces. They
begin by choosing a given set of functions that will be 
integrable. The functions chosen are ``cylindrical functions''.
These are functionals on the infinite dimensional space that
really only depend on certain ``directions'' or ``projections''
of the space against a set of Schwarz test functions. The
projection is achieved through the use of Wilson loops, or
even more easily, spin networks. It might appear that 
cylindrical functions are too simple to be able to capture
interesting physics. But for the case of a scalar field,
for instance, the Fock measure can be constructed with
cylindrical functions. In the case of spin networks the
resulting measure of integration is really simple, it 
just states that spin networks are actually an orthogonal
basis, ie, $<s_1|s_2>=\delta_{s_1,s_2}$ where the 
delta on the right hand side is one if the spin networks
are equal or zero otherwise. The measures constructed are
naturally diffeomorphism invariant. If one considers 
the class of spin networks related by diffeomorphisms with
$s_1$ and the class related with $s_2$ one can construct
an inner product on the classes simply by demanding that
their inner product be zero if no member of the $s_1$
class coincides with at least one member of the $s_2$ class.
The subject of measures of integration has several 
mathematical subtleties. Physicists can get a very readable
succinct account in the paper by Ashtekar, 
Marolf and Mourao \cite{Ashtekar:et}.

\subsection{Areas and volumes}

Most quantities of physical interest will involve products of
the fundamental fields and therefore will require regularization.
The latter is a non-trivial subject in quantum gravity. 
Most regularization procedures used in particle physics require
the use of metric information. In particle physics the metric
is a c-number. But this is not the case in gravity. If one 
insists in using regularization procedures that involve the
metric, one should consider it an operator. This can complicate
quite a bit the task of regularizing expressions. Alternatively,
one can introduce an external c-number metric into the formalism,
and hope that after one regularizes, there is no trace left of 
this artificial element in the construction. It is of
interest to notice that some operators of (limited) physical
interest can indeed be computed that are well defined in 
spite of the use of regularizations. These operators represent
the area of a surface and the volume of a region of space.
At first it appears that these operators will be difficult
to regularize. The classical expression for the area of a 
surface $\Sigma$ is $A(\Sigma)=\int \sqrt{\tilde{E}^a_i
\tilde{E}^b_i n_a n_b} d^2x$ where $n_a$ is the 
normal to the surface. The presence of the square root 
might at first suggest that regularization will be 
problematic. However, partitioning the area in small
elements of area one can quickly see that for the quantity
inside the square root spin network states are actually
eigenstates. The end result for main portion of the spectrum of the
area operator is
\begin{equation}
\hat{A}(\Sigma) |s> =\sum_L \sqrt{J_L (J_L+1)} \ell_P^2 |s>
\end{equation}
where the sum is over all links of the spin network that 
pierce the area and $J_L$ is the valence of the link $L$.
We see that areas are quantized in terms of the Planck
length squared $\ell_P^2$. This was first noticed by 
Rovelli and Smolin \cite{Rovelli:1994ge}. Later Ashtekar and Lewandowski 
\cite{Ashtekar:1996eg} did
a comprehensive analysis of the spectrum of the area 
operator.  The quantization
of the area reveals another surprise. Most people expected
areas to be quantized, but the expectation was that the
spectrum would be equally spaced, i.e, $n \ell_P^2$. 
It is not. This has consequences. Bekenstein and Mukhanov
\cite{Bekenstein:1995ju}
have shown that assuming that the spectrum of the area is
equally spaced has serious implications for the validity 
of the thermal spectra of black holes. Rovelli and collaborators
\cite{Barreira:1996dt} have shown
that these problems can be solved by considering the 
correct spectrum.

For the volume operator results are quite similar. The
volume operator acting on a spin network gives a nonzero
result if within the region considered there are 
intersections of valence equal or larger than four. One 
gets a picture of spin networks in which each links carries
``quanta of area'' and each intersection of valence four
or larger carries ``quanta of volume''. Both operators are
finite and well defined without reference to any background
metric structure in spite of the fact that they
had to be regularized. Several calculations of the spectrum
of the volume can be found in \cite{volume}.

\section{Physical predictions: Gamma ray bursts}

It is clear that one cannot really discuss any physics
emerging from quantum gravity until one has dealt with
the Hamiltonian constraint. Attempting to do so would be
equivalent to trying to do physics after handling two 
of the three components of Gauss' law in $SU(2)$ Yang--Mills 
theories. One can attempt to do some calculations ``at
the kinematical level'' (i.e. ignoring the constraints) in the
hope that some of the basic features of the calculations will 
persist when the constraints are enforced, but this is not
guaranteed. It is important to preface the discussion of this
section with these caveats since they very much apply to what
we will discuss.

It took everyone by surprise when Amelino-Camelia and collaborators
\cite{Amelino-Camelia:1997gz}
at CERN argued that in the detection of gamma ray bursts one 
could find traces of quantum gravity phenomenology. For years 
it had been common lore that quantum gravity required energies
so high that it could only have relevant effects in the big
bang or inside black holes. The possibility of detecting 
quantum gravity effects via gamma ray bursts goes as follows:
the gamma rays that arrive on Earth have travelled a very 
long distance, since gamma ray bursts are expected to be 
cosmological. That distance appears even larger if one
measures in terms of the number of wavelengths of a 
gamma ray. If one assumes that when a wave propagates through
the ``quantum foam'' each wavelength gets disturbed by an
amount of the order of the Planck length, then the smallness
of this number can be compensated by the huge number of 
wavelengths involved in traveling from the burst to Earth.
If one inputs numbers it turns out that detectable dispersions
(differences in times of arrival) of $0.01$ seconds in gamma 
rays that differ by $300keV$ as those detected by the BATSE
experiment imply that quantum gravity has to happen at 
energies larger than $10^{16}GeV$ in order not to be visible. 
This is only three orders of magnitude away from the Planck
scale! This led to a lot of interest in these observations.

Within loop quantum gravity some calculations have been performed to
attempt to estimate these effects, at a kinematical level
\cite{Gambini:1998it}. The calculations require a number of
simplificatory assumptions. Otherwise one would have to deal with
Einstein-Maxwell theory and work out a semiclassical limit. In general
the assumptions have been that the electromagnetic field has been
treated classically and only the Maxwell part of the Hamiltonian
constraint is considered. One finds modified Maxwell equations that
imply that there is birefringence in the propagation of waves. Similar
results can be found for neutrinos \cite{Alfaro:1999wd}. 
The birefringence for photons has
been severely constrained experimentally \cite{Gleiser:2001rm}. 
A much more careful
recalculation of the effects done recently confirms several of
the general features of the original calculations 
\cite{Thiemann:2002vj}.

The main problem with these predictions is that in order
to have a non-vanishing effect at the lower order in terms
of the energy of the gamma rays, one needs to introduce
rather unnatural assumptions in terms of the quantum state
considered (otherwise one could not generate a birefringence,
which is tantamount to a parity violation). If one does not
make these assumptions, then the effects only arise at the
next order in $E/E_{Planck}$ and are completely undetectable.
In terms of the original work of Amelino Camelia et al.
they postulated a non-standard dispersion relation of the
form 
\begin{equation}
c p^2 = E^2 (1+\alpha {E\over E_{Planck}}+O\left({E^2\over E_{Planck}^2}\right)
\end{equation}
and the effects would be observable if $\alpha$ is non-vanishing.
A non-vanishing $\alpha$ implies a fractional power in a dispersion
relation, which is unusual.

More importantly, all these calculations are implying that one
is violating Lorentz invariance. This is a huge step to take.
There is significant discussion of the implications in the 
current literature (see \cite{Magueijo:2002am} and references
therein). 

\section{Thiemann's Hamiltonian constraint}

One of the initial encouragements that the new variables
introduced was that the Hamiltonian constraint appears as
a polynomial function of the fundamental variables. This
suggested that one could perhaps promote it to a quantum
operator and several attempts to regularize it were 
carried out. However, there is an obvious fundamental
flaw in attempting this. The Hamiltonian constraint is
quadratic in the triads. The triads are densities of 
weight one, meaning that the constraint is a density of
weight two. More precisely, the version of the constraint
that is nice and polynomial is a density of weight two.
One could turn it into a density of weight one by 
dividing by the determinant of the metric, but then the
resulting operator would be complicated and non-polynomial.
Why is it a problem that it is a density of weight two?
Suppose we wished to promote it to an operator in the
loop or spin network representation. What could such
an operator be? We have at our disposal a manifold,
and a set of lines in it. We have available a density
of weight one, the Dirac delta, which is naturally
defined on any manifold. But we do not have a density
of weight two. And we cannot multiply Dirac deltas.
Therefore if one found a regularization of the 
doubly densitized Hamiltonian constraint, what has
to be happening is that one provided the extra density
weight via the regulator. And therefore the imprint
of the regulator will not disappear upon regularization.

All these difficulties were bridged when Thiemann
\cite{Thiemann:1996aw}
 discovered
how to handle the single-densitized Hamiltonian 
constraint. The expression for the constraint is,
\begin{equation}
\tilde{H} = {\tilde{E}^a_i \tilde{E}^b_j F_{ab}^k \epsilon^{ijk} 
\over \sqrt{\tilde{E}^a_i  \tilde{E}^b_j\tilde{E}^c_k \epsilon^{ijk}\epsilon_{abc}}}, 
\end{equation}
and Thiemann noticed that
\begin{equation}
{\tilde{E}^a_i \tilde{E}^b_j  \epsilon^{ijk} 
\over \sqrt{\tilde{E}^a_i  \tilde{E}^b_j\tilde{E}^c_k 
\epsilon^{ijk}\epsilon_{abc}}} =
2 \left\{A_a,V\right\}
\label{thiemid}
\end{equation}
where $V$ is the volume of the three manifold. The Hamiltonian constraint can
therefore be written as,
\begin{equation}
H(N)=\int d^3x N(x) {\rm Tr}(F_{ab} \{A_c,V\}) \epsilon^{abc}.
\end{equation}

When we first discussed the Hamiltonian constraint with the new variables,
we noted that it was important that we take the Immirzi parameter to be the
imaginary unit. That made certain non-polynomial terms disappear, but at the
price of making the variables complex. Thiemann noted that through a similar
use of identities as the one we discussed, these non-polynomial terms could
also be reexpressed in terms of Poisson brackets. Therefore there is no
need anymore of taking the Immirzi parameter to be imaginary and from now on
one can work with variables that are completely real.

Thiemann proposed a quantization for the above mentioned Hamiltonian constraint.
The procedure consists in introducing a lattice. He chooses an irregular 
lattice (tetrahedral). In terms of this lattice, he approximates the expression
for the classical Hamiltonian constraint using holonomies. Omitting many
details, the idea is that the ``$F_{ab}$'' term is represented by a closed
loop going around a triangle on one of the faces of the elementary tetrahedron
and the ``$A_c$'' is represented by a line holonomy that is retraced to recover
gauge invariance. The classical Hamiltonian constraint discretized on the lattice
is therefore only a function of holonomies and the volume of the manifold. The 
attractive aspect of this is that both holonomies and the volume of the manifold
can be represented by well defined finite operators in the spin network 
representation. Therefore producing a well defined, finite Hamiltonian constraint
is tantamount to ``putting hats on the classical expression'' since all the
ingredients can be naturally quantized without divergences! 

There are a couple of caveats that need to be noted. The ``$F_{ab}$'' can be
constructed by many different kinds of elementary loops. As long as they 
shrink to a point when the lattice is refined they all will represent properly 
the curvature. This indicates that there is therefore
huge ambiguity in how to define the operator. An additional ambiguity is the
valence of the holonomy that represents the curvature 
\cite{Gaul:2000ba}. Moreover, a crucial
element for the Hamiltonian to be well defined is that it act on diffeomorphism
invariant states. On such states the details of how the holonomy that represents
the curvature is placed with respect to the spin network are immaterial. 
This, in turn, ensures that the resulting quantum theory is consistent. 
If one acts with two Hamiltonian constraints, the two loops added are indistinguishable
from each other and therefore if one acts in the opposite order the final result
is the same. The Hamiltonian constraint therefore commutes with itself.
Now, the classical Poisson algebra of constraints stated that the Poisson bracket
of two Hamiltonians should be proportional to a diffeomorphism. If one
promotes this to a quantum operatorial expression and acts on diffeomorphism
invariant states, the right hand side will give zero since they are annihilated
by diffeomorphisms. Therefore the commutator of two Hamiltonians should vanish
as well \cite{Thiemann:1997rv}.

Thiemann goes on to show that similar constructions can be carried out
for general relativity coupled to fundamental matter fields:
Yang--Mills, Higgs, Fermions \cite{Thiemann:1997rt}. This achievement
is quite remarkable. We are in the presence of the first finite, well
defined, anomaly-free, non-trivial theory of quantum gravity ever
presented. The theory fulfills the promise of acting as a ``natural
regulator of matter fields'' in the sense that no divergences are
present when the theory is coupled to matter.

Is quantum gravity finally achieved? The answer is still not known. What has
been found is a theory (more precisely infinitely many theories due to the
ambiguities) that are well defined. Although this is no small feat in this
context, We do not know if any of these theories contains the correct physics
of gravity. This will only be confirmed or contradicted when a semiclassical
approximation is worked out so we can make contact with more familiar results.
Active investigations along these lines are being pursued by Thiemann and
collaborators \cite{Thiemann:2002vj} and Ashtekar and collaborators
\cite{Ashtekar:2001xp}.

There are some aspects of Thiemann's construction that appear somewhat
troubling. The same construction can be worked out in $2+1$
dimensional gravity \cite{Thiemann:1997ru}. If one studies the
solutions of the quantum Hamiltonian constraint one finds many more
states than the ones allowed by Witten's theory.  However, if one
demands that they be normalized with the inner product we discussed,
the Witten sector is all that is left. This can be seen as a positive
result (after all, we get the correct theory) or as a negative one
(the correct theory only is recovered after choosing carefully an
inner product). It appears that in $3+1$ dimensions Thiemann's
Hamiltonian also admits too many solutions. The fact that the
constraint algebra can only be recovered on diffeomorphism invariant
states, where it is only Abelian, is also troubling. Though again,
there is no genuine interest in states that are not diffeomorphism
invariant. Other worries were expressed in a paper by Smolin
\cite{Smolin:1996fz}. The
general consensus at the moment is that there appear to be worries
that the Hamiltonian does not capture the correct physics, but no one
can make a theorem out of the worries to prove that Thiemann's
Hamiltonian is wrong. The verdict will come when further explorations
of the semiclassical approximation are worked out.

\section{Bojowald's cosmologies}

The idea of exploring quantum gravity effects in the simplified
context of cosmological models has held appeal over the years. Yet,
due to the lack of a theory of quantum gravity, the approach taken was
rather bizarre. People would consider general relativity, then reduce
the classical theory to only cosmological metrics (which in the case
of homogeneous cosmologies reduces the equations to ordinary
differential equations, losing the field theoretical nature of general
relativity).  The resulting theory was then quantized and some
interpretations were attempted. The main criticism that was levied
against this kind of investigations is that ``imposing a symmetry then
quantizing'' does not have to agree with ``a sector of the quantum
theory with a given symmetry''.  That is, there is no guarantee that
what one sees in quantum cosmology will appear at all when one gets a
handle of the full theory and studies cosmological situations.

With Thiemann's introduction of viable theories for quantum gravity,
it therefore became ubiquitous to attempt to study quantum cosmology
``properly''. That is, study the sector of the full quantum theory of
gravity that approximates homogeneous cosmologies. This is what
Bojowald \cite{Bojowald:1999tr} set out to do. He finds that in
isotropic and homogeneous 
quantum cosmology states reduce to ``spin networks with only one
link''. This is understandable since everything happens ``at a single
point'' in a homogeneous model. The presence of the link is needed to
make sense of the operators involving connections (one needs more than
a point to have a notion of a connection!).  The quantum states
therefore are labeled by an integer $|n>$ which corresponds to the
valence of the single link of the spin network.  Bojowald constructs a
version of Thiemann's Hamiltonian acting on these states. He also
finds a well defined version of the volume operator.

One of the long held beliefs in quantum cosmology is that quantum
effects will eliminate the big bang singularity. In Bojowald's case
this is actually realized in practice. Considering the case of a flat
Robertson--Walker metric $ds^2=-dt^2+a(t)^2 dx^2$, he finds that he
can find a finite, well defined expression for the operator
$1/a(t)$. This is done through similar identities as those that led to
equation (\ref{thiemid}). Since the resulting operator is finite, it 
suggests
that the singularity can be avoided. Remarkably, if one analyzes the
relationship of $a(t)$ with the volume operator (which should be
$1/a(t)=V^{-1/3}$ such relationship does hold quantum mechanically
when the universe is ``large''. But when the universe becomes of the
size of a few Planck volumes, the relationship is broken, the volume
goes to zero but $1/a$ remains finite \cite{Bojowald:2001xe}. The
avoidance of the singularity can be implemented concretely in this
approach through discrete equations of motion that actually never
become singular. And the theory can be coupled to various matter sources
without introducing singular behaviors through the use of the defined
$1/a(t)$ to implement the couplings.

Bojowald goes on to introduce a notion of time for these
cosmologies. Since everything is discrete, his notion of time is
discrete too. The evolution equations are recursion relations and he
shows that for large universes they reproduce the results of usual
Wheeler-DeWitt-based quantum cosmology
\cite{Bojowald:2001ep}. And these evolution equations also allow to
evolve non-singularly through the point where one expects the 
singularity classically. Remarkably, even an argument for inflation
being generated by quantum gravity can be found in this context
\cite{Bojowald:2002nz}.

The fact that the cosmological reduction of Thiemann's Hamiltonian
appears to give the correct physics of quantum cosmology is considered
by some as an indication that the right physics of gravity is contained.
One should be aware of the fact, however, that Bojowald's construction
implies a limiting procedure. Quantum states peaked on homogeneous
cosmologies are really distributions and therefore one needs to extend
the operators defined for other states to them. This extension is non-trivial
and there might be ambiguities in it that allow to ``correct'' things in
order to get the right physics. Although this is not what Bojowald set
out to do, it might have accidentally happened. Moreover, part of the
worries about Thiemann's Hamiltonian have to do with the constraint
algebra. In homogeneous cosmologies, since everything takes place
``at a point'' there is only one Hamiltonian constraint that therefore
is obviously Abelian, which agrees with Thiemann's general result.

Nevertheless, it is striking that detailed attractive predictions in the
cosmological context can be extracted from the proposed Hamiltonian
constraint.

\section{Consistent discretizations: a new framework?}

When we discussed Thiemann's Hamiltonian constraint we mentioned that he
started from a given classical theory in the continuum and introduced
a lattice to discretize the Hamiltonian constraint. The lattice Hamiltonian
is then promoted to a quantum operator naturally. The idea of using lattices
to regularize gravity is not new. The novelty is the use of the recently
acquired knowledge about well defined operators and states. Lattice approaches,
however, are plagued by difficulties to which Thiemann's approach may not
be immune. The difficulty has to do with the fact that in the case of 
general relativity lattice regularizations breaks the symmetry of the
theory under diffeomorphisms (on the contrary, in the Yang--Mills case,
lattice gauge theory has the advantage of providing a gauge invariant
regularization). The theories one gets on the lattice therefore have
a considerably different structure than the theory they attempt to
approximate in the continuum. It is the personal impression of the
author that at the time of quantizing the discrete theories, one needs
to take their structure seriously.

In particular, most discrete approximations that one constructs for general
relativity, end up being inconsistent (their equations do not admit
a single solution). This is well known, for instance, in numerical
relativity. When one wishes to integrate the Einstein equations on a
computer, they are approximated by finite difference equations.
Whereas in the continuum theory if one solves the constraint equations
initially, the evolution equations guarantee the constraints will
hold at all time, this is not true of the discrete equations.
There is therefore no way to satisfy simultaneously the constraint
equations (at all times) and the evolution equations. Most people in
numerical relativity use ``free evolution'' that is, they accept
that they will fail to satisfy the constraints at later times as
part of the numerical error of the solution they incur.

In quantization the last argument does not work. If a theory is 
inconsistent, there is little sense in attempting a quantization.
Most attempts to lattice quantum gravity suffer from this problem.
For instance, when one discretizes the Hamiltonian constraint, the
discrete constraints fail to close an algebra (this is reasonable,
since algebras are associated with infinitesimal symmetries and
nothing can be infinitesimal in a discrete theory). If the 
constraints do not close an algebra one can generate further 
constraints by taking Poisson brackets. If one is not careful one
ends up with too many constraints and the theory has no solutions.
This is the canonical manifestation of the inconsistency.

These kind of problems are very basic and are even present in 
very simple systems. For instance, if one considers a Newtonian
particle and discretizes Newton's equations, it is a well known
fact that energy fails to be conserved. In fact astronomers who
wish to follow planetary motion have known this for a long time
and construct special discretizations of Newton's laws that
automatically conserve energy and angular momentum. Our goal is
to find something similar in the gravitational context, that is,
a discretization scheme that automatically preserves the 
constraints.

T.D. Lee \cite{TDLee} proposed a way to fix the problems of the
Newtonian particle that can be easily translated to the gravitational
case.  The idea consists in enforcing the constraints through a
suitable choice of Lagrange multipliers. In the case of general
relativity, one chooses the lapse and shift in such a way that in the
next step the constraints are satisfied.  This has no counterpart in
the continuum theory. One has four constraints to enforce and four
quantities to solve for (the equations to be solved are coupled
algebraic non-linear equations).

We have recently worked out \cite{consistent} the canonical theory for
such consistent discretizations and applied it to Yang--Mills and BF
theory and presented the prescription for the gravitational case. In
this approach, since the constraints are automatically satisfied, most
of the conceptually hard problems that arose due to attempting to
impose the constraints disappear.  Quantum gravity become a
conceptually clear yet computationally challenging problem. There
are many attractive features in this approach. Since the initial data
fixes the lapse and the shift, and quantum mechanically one
generically has a superposition of initial data, one automatically has
a superposition of discretizations. The goal of ``averaging out'' over
all discretizations is implemented naturally. We have applied these
ideas to a cosmological model. Classically one finds that if one runs
the model backwards, unless one fine tunes the initial data, no
singularity is present. This is natural, generically the singular
point will not fall on the lattice.  When one quantizes however, the
fact that the singularity classically only occurs for a set of measure
zero in the possible initial data implies that the singularity is not
present. We see a remarkable agreement with Bojowald's prediction,
although the details and motivation are different.

Much more will have to be explored to see if the consistent
discretizations are a viable route for quantization. In particular
we have little experience with the complicated non-linear equations
that fix the lapse and the shift. What if they quickly generate
negative lapses or singularities? It is imperative that experience
be gained, particularly in midi-superspace examples where there
are field theoretic degrees of freedom. Even these cases are
computationally challenging given the complexity of the algebraic
non-linear coupled equations that determine the lapse and the shift.

\section{Summary}

The last 18 years have seen a renaissance of canonical quantum
gravity.  The field has been brought to a complete new level in terms
of mathematical sophistication and possibilities for discussing
physical consequences and applications. It is becoming more evident by
the day that not only it is not clear that general relativity has a
problem at the time of its quantization, but that actually the
quantization of Einstein's theory has a lot to teach us about
physics. This physics might be of interest just in the context of pure
general relativity or in the context of proposed unified theories of
all interactions, like string theory

\section{Acknowledgments}
I am grateful to A. P\'erez, M. Bojowald, J. Lewandowski, T. Thiemann
for comments on the manuscript.  I also wish to thank the organizers
of the Brazilian school for the invitation to participate. This paper
was completed at the Erwin Schr\"odinger Institute in Vienna. My work
in this field has for many years been in collaboration with Rodolfo
Gambini. This work was supported by grants NSF-PHY0090091, funds from
the Horace Hearne Jr. Institute for Theoretical Physics and the
Fulbright Commission in Montevideo.

\end{document}